\documentclass[final,5p,times,twocolumn,number]{elsarticle}

\usepackage[english]{babel}
\usepackage{epstopdf}
\usepackage{multiobjective}

\usepackage[version=0.96]{pgf}
\usepackage{tikz}
\usepackage{tikzscale}
\usepackage{tikzinclude}
\usepackage{epstopdf}
\usetikzlibrary{arrows,shapes,snakes,automata,backgrounds,petri}
\usepackage{hyperref}
\usepackage{multirow}

\graphicspath{{figures/}}
\usepackage{epstopdf}

\usepackage{enumitem}
\usepackage{caption}
\usepackage{subcaption}

\usepackage{float}

\usepackage[colorinlistoftodos]{todonotes}

\usepackage{xcolor,colortbl}

\usepackage{amssymb}
\usepackage{amsmath}

\journal{Decision Support Systems}
\renewcommand{\cite}[1]{\citep{#1}}

\usepackage{etoolbox}
\patchcmd{\pprintMaketitle}
 {\ifvoid\absbox\else\unvbox\absbox\par\vskip10pt\fi}
 {\ifvoid\absbox\else\clearpage\unvbox\absbox\par\vskip30pt\fi}
 {}{}
\patchcmd{\pprintMaketitlee}
 {\hrule\vskip12pt}
 {}
 {}{}
\patchcmd{\pprintMaketitle}
 {\hrule\vskip12pt}
 {}
 {}{}
\appto{\pprintMaketitle}{\clearpage}

\begin{document}
\begin{frontmatter}


\title{Google COVID-19 community mobility reports: insights from multi-criteria decision making}                      




\author[1]{Gabriela Cavalcante~da~Silva} 
\ead{gabrielacavalcante@ufrn.edu.br}


\address[1]{IMD, Universidade Federal do Rio Grande do Norte (UFRN), Brazil}

\author[2]{Sabrina Oliveira} 
\ead{oliveira.sabrina@gmail.com}


\address[2]{Postgraduate Program in Mathematical and Computational Modeling, CEFET-MG, Brazil}

\author[3,4]{Elizabeth~F. Wanner} 
\ead{efwanner@cefetmg.br}


\address[3]{Computer Engineering Dept., CEFET-MG, Brazil}
\address[4]{Computer Science Group, Aston University, UK}

\author[1]{Leonardo~C.~T. Bezerra}
\ead{leobezerra@imd.ufrn.br}


\cortext[cor1]{Corresponding author: Gabriela Cavalcante da Silva,  IMD, Universidade Federal do Rio Grande do Norte (UFRN), Brazil,  Av. Cap. Mor Gouveia, 3000 - Lagoa Nova, Natal - RN, CEP 59078-970, Brazil}

\nonumnote{Declarations of interest: none}


\begin{abstract}
Social distancing (SD) has been critical in the fight against the novel coronavirus disease (COVID-19). To aid SD monitoring, many technology companies have made available mobility data, the most prominent example being the community mobility reports (CMR) provided by Google. Given the wide range of research fields that have been drawing insights from CMR data, there has been a rising concern for methodological discussion on how to use them. Indeed, Google recently released their own guidelines, concerning the nature of the place categories and the need for calibrating regional values. In this work, we discuss how measures developed in the field of multi-criteria decision making (MCDM) might benefit researchers analyzing this data. Concretely, we discuss how Pareto dominance and performance measures adopted in MCDM enable the mobility evaluation for (i) multiple categories for a given time period and (ii) multiple categories over multiple time periods. We empirically demonstrate these approaches conducting both a region- and country-level analysis, comparing some of the most relevant outbreak examples from different continents. 


\end{abstract}










\end{frontmatter}

\newcommand{\LEO}[1]{\footnote{LEO: #1}}

\section{Introduction}
\label{sec:intro}

In December 2019, the World Health Organization~(WHO) Country Office in the People’s Republic of China reported cases of pneumonia of unknown etiology.
\footnote{\url{https://who.int/news-room/detail/29-06-2020-covidtimeline}}
In January 2020, WHO named \textit{SARS-CoV-2} the novel coronavirus responsible for these cases, and the acute respiratory syndrome it caused \text{COVID-19}~(\textit{coronavirus disease 2019}). Still in January, WHO classified COVID-19 as a public health emergency of international concern. In March, COVID-19 had cases reported from all continents, and WHO declared it a pandemic.\sloppy

The rapid spread of COVID-19 and the deaths it produced triggered a global scientific rush. Indeed, by April over 20 surveys on COVID-19 could already be identified~\cite{Yu2020assessment}. However, this fast publication pace raises concerns on the quality of the works produced. Specifically, many such works have not undergone a peer reviewing process, casting uncertainty as to the methodological decisions adopted by their proposers. This is even more important in the COVID-19 context, where knowledge from complementary research fields is required to propose multi-disciplinary solutions to fight the pandemic.

Among the most relevant topics in this multi-disciplinary COVID-19 research context is social distancing~(SD), which WHO actively promotes as a non-pharmaceutical intervention against COVID-19~\cite{WHO-sd}.
To aid  SD monitoring, information technology~(IT) companies have been publishing anonymized mobile device location history data, among which Google~\cite{Aktay2020cmr}. In particular, the community mobility reports (CMR) provided by Google comprise over 130 countries, some of which further detailed on a regional level. More importantly, the data for a given country/region is a collection of time series for six place categories. Therefore, SD analysis based on CMR data requires theoretical approaches from the fields of time series analysis~(TSA) and multi-criteria decision making~(MCDM)~\cite{Zitzler2003performance,Goldberg1989genetic,Laumanns2002epsilon}. Furthermore, the existence of six different place types renders this a \textit{many-criteria} decision making problem~\cite{Li2015many}, a particularly challenging specialization of MCDM. Indeed, Google itself released guidelines to aid this analysis~\cite{CMR}, such its awareness of the many different research fields it bridges.

The goal of this paper is to discuss methodological approaches that can help analyze SD based on CMR data, with a special focus on MCDM. In principle, the temporal nature of this data requires techniques from TSA, from which we comment on the effects of (i)~alternatives to reduce seasonality effects~\cite{Cleveland1990stl}; (ii)~techniques to aggregate temporal dynamics, and; (iii)~different granularities for temporal discretization. Next, we discuss the effects of different MCDM approaches that help provide a multi-criteria perspective to this analysis, namely (i)~Pareto dominance~\cite{Zitzler2003performance} and (ii)~Pareto-related measures that help deepen the analysis~\cite{Goldberg1989genetic,Laumanns2002epsilon}. Moreover, we adopt visualization techniques that help understand the MCDM conclusions even in this many-criteria context.

To empirically demonstrate the approaches discussed in this work, we conduct both a region- and a country-level analyses comparing some of the most relevant outbreak examples from several continents. Specifically, the region-level analysis comprises locations that experienced some of the most severe pandemic outbreaks~\cite{jhu}, namely Lombardia~(Italy), Île-de-France~(France), New York~(United States), and São Paulo and Amazonas~(Brazil). By its turn, the country-level analysis comprises the Americas~(Argentina and Canada), Europe~(Germany and Spain), Asia~(Japan and South Korea), and Oceania~(New Zealand). 

Insights observed confirm the effects and importance of the methodological resources discussed in this work. Some of these insights were already expected, such as the effects of calibration and scaling on the comparison, and the need for seasonality approaches to process the original raw data. Other insights are specific to CMR assessment, such as (i)~how the nature of the different place categories interact with the experimental factors considered and (ii)~how different temporal granularities reveal contrasting dynamics among localities. More importantly, we observe how an MCDM perspective provides nuances that scalarized approaches would conceal, and assist the decision-maker both visually and when configuring tolerance levels. Concerning localities, the region-level analysis indicates that the European regions considered were able to better promote social distancing than the American ones, though a fine-grained temporal assessment indicates that this was only achieved after a few weeks of mobility restriction. 

Below, we summarize the main contributions of this paper:
\begin{itemize}
    \item A theoretical discussion on methodological alternatives for a multi-criteria temporal analysis of social distancing through mobility data.
    \item An empirical mobility reduction analysis of some of the most relevant outbreak examples from several different continents in a region- and country-level.
\end{itemize}

The remainder of this paper is organized as follows. We initially briefly review related work in Section~\ref{sec:related-work}. Next, Section~\ref{sec:methodology} discusses the methodology we adopt in this work to enable a multi-criteria social distancing analysis from mobility data. The country- and region-level analyses are detailed in Section~\ref{sec:results}. Finally, we conclude and discuss future work in Section~\ref{sec:conclusion}.

\section{Related work}
\label{sec:related-work}

As previously discussed, the scientific rush stirred by the COVID-19 pandemic has led to a significant number of works that have been made available online not having undergone peer review. In this section, we consider only works that have already been peer reviewed, and group them according to how they relate to this work. Specifically, we first discuss works that relate social distancing and COVID-19, and how mobility data can be used in this context. Later, we review multi-criteria decision making approaches that have been employed in the COVID-19 scenario.

\paragraph{Social distancing~(SD)}
WHO has appointed SD as the main non-pharmaceutical intervention against \text{COVID-19}~\cite{WHO-sd}. Besides confirming this effectiveness, \citet{10.1093/cid/ciaa889} have indicated that SD may also be related to easing the symptoms patients develop. Yet, SD adherence is subject to cultural traits~\cite{HUYNH2020104872}, besides socioeconomical factors. For instance, Asian countries are more used to facial masks, and developing countries make less use of remote work.
In this context, a systematic quantification and analysis of the impact of SD policies on people mobility may better inform health decision-making strategies during the pandemic. 

\paragraph{Mobility data}
Early on, the relevance of SD monitoring through mobility data was highlighted by~\citet{Buckeeetal2020}, who also discussed the need for anonymity through aggregation and data privacy protocols. While local governments have often relied on sensitive triangulation data from telecommunication companies, some IT companies have anonymized their mobility data and published them online~\cite{CMR,apple,inloco}. Given the fast-changing nature of this data, many works that monitor SD using mobility data have been made available not having undergone peer review. An exception that is closely related to the assessment we conduct in this work used data from a Brazilian IT company~\cite{inloco} to evaluate SD in its home state~\cite{ENDO2020}.  In addition, a few works use mobility data to retrospectively model \text{COVID-19} spread in different locations~\cite{PDE_covid,Kraemeretal2020}.

\paragraph{Multi-criteria decision making}
As previously discussed, SD policies employed to contain the dissemination of COVID-19 must also be economically sustainable, allowing for planned social movement. This conflictual search for balance has been modeled as lockdown/lifting policies and studied from a multi-criteria perspective in different works~\cite{10.3389/fpubh.2020.00294,SANGIORGIO2020}. Another application of multi-criteria decision making techniques in the context of COVID-19 is model evaluation. For instance, \citet{covid_19_DM_preventive} have evaluated the importance of recommended protocols for COVID-19 prevention, whereas \citet{topis_covid19} have evaluated predictive diagnostic models from a multi-criteria perspective. Yet, we have not identified works that assess mobility data from an MCDM perspective.
\begin{figure*}
\includegraphics[width=\textwidth]{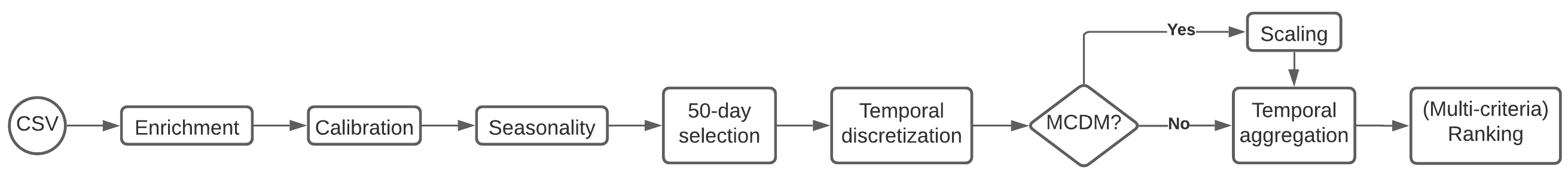}
\caption{Methodology discussed in this work, comprising data preparation, time series analysis, and multi-criteria decision making.}
\label{FIG:methodology}
\end{figure*}

\section{Methodology}
\label{sec:methodology}

As previously discussed, the mobility data we assess in this work is provided by Google as community mobility reports~(CMR). In this section, we initially detail CMR data and how we enrich it with mobility restriction-related dates and prepare it for analysis. Next, we discuss time series analysis concepts and techniques, critical to this assessment as the CMR data is provided as a collection of time series. Finally, we discuss concepts and techniques from multi-criteria decision making~(MCDM) which allow us to assess the CMR data from a multi-criteria perspective.  The whole process described in this section is summarized in Figure~\ref{FIG:methodology}.

\subsection{Data acquisition, description, and enrichment}

CMRs are provided by Google weekly as printed document files~(PDFs), first created on March 2020. Additionally, Google makes CMR data available as a comma-separated values~(CSV) file, which comprises data from all reports ever produced. In total, CMR data comprises 135 countries, some of which are further detailed on a regional level. Data is collected from users who willingly enable their location history, and is anonymized as described in~\citet{Aktay2020cmr}.

Data for each country is provided as a collection of six time series, one for each place category created by Google, given in Table~\ref{tab:place_categories}. Each time series starts in February 15$^\text{th}$, 2020, and currently extends until August 16$^\text{th}$, 2020. For a single timestamp and category, the given value is computed relatively to a baseline, namely the median value, for the corresponding day of the week, computed for the period between January 3$^\text{rd}$, 2020 and February 6$^\text{th}$, 2020~\cite{CMR}.

\begin{table}[!t]
    \centering
    \caption{Place categories and descriptions from Google CMR~\cite{CMR}.}
    \label{tab:place_categories}
    \scalebox{0.8}{
    \begin{tabular}{p{3.3cm}|p{6.7cm}}
        \hline
        \textbf{Category} & \textbf{Places} \\ 
        \hline
        \textit{Grocery}\linebreak \& \textit{pharmacy} & Grocery markets, food warehouses, farmers markets, specialty shops, drug stores, and pharmacies\\ 
        \hline
        \textit{Parks} & Local parks, national parks, public beaches, marinas, dog parks, plazas, and public gardens\\
        \hline
        \textit{Transit stations} & Public transport hubs, e.g. subway and bus stations\\
        \hline
        \textit{Retail}\linebreak \& \textit{recreation} & Restaurants, cafes, shopping centers, theme parks, museums, libraries, and movie theaters\\
        \hline
        \textit{Residential} & Places of residence\\
        \hline
        \textit{Workplaces} & Places of work\\
        \hline
    \end{tabular}
    }
\end{table}
For the countries and regions we adopt in this analysis, we have further enriched the data with (i)~their initial mobility restriction dates and (ii)~the dates when these restrictions started to be relaxed. Specifically, we have used school suspension dates as the initial restriction dates, as our preliminary assessment showed this was the restriction measure that most significantly affected mobility. 

\subsection{Data preparation}

As previously discussed, Google has recently provided guidelines for the assessment of its CMR data~\cite{CMR}. Among the most relevant recommendations are (i)~the need for calibrating data in a locality-wise basis; (ii)~handling noise incurred by holidays or other exceptional circumstances, and; (iii)~the difference in magnitude between categories. Here, we discuss how we handle the first recommendation, and in the following sections we discuss the remaining ones.

The need for calibrating data for each locality comes from the baseline period adopted by Google when computing relative values. More precisely, the assumption that the first five weeks of the year represent a period of average mobility is very strong, specially in a global scenario as we investigate in this work. Indeed, for most localities considered in this analysis, February mobility presented significant deviations from the average mobility that the baseline period should represent. A typical approach in machine learning when comparing data from different sources is \textit{standardization}, i.e., the data from each source is first centered such that its mean is set to zero, and then scaled such that its standard deviation equals one. 

In this work, for each locality and category we process the whole time series to ensure that the data previous to the first mobility restriction date presents zero mean. Our goal is to promote comparability between different localities, while preserving data interpretability. More precisely, our mean-centering procedure aims to ensure that the data prior to mobility restriction measures represent the average mobility for that locality and category.%
\footnote{We remark that the magnitude of the deviations seen for February data for most localities is such that ideally it should be discarded. Yet, this was not an option in this investigation as we include Lombardia, for which the first mobility restriction measure dates of February.}
If, instead, we had centered the data to ensure the whole time series presented zero mean, we would have compromised the interpretability of the series after the restriction dates. To also preserve interpretability, we have chosen not to scale the data at this point, but later in the MCDM analysis. 

\subsection{Time series analysis~(TSA)}

Given the temporal nature of the CMR data, we discuss three sets of concepts and techniques from TSA which we employ in our assessment: (i)~seasonality effect reduction; (ii)~time discretization granularity, and; (iii)~temporal dynamics aggregation. For each set, we highlight in boldface the experimental options we adopt, summarized in Table~\ref{tab:factors_levels}. 

\begin{table}[!t]
    \caption{Experimental factors considered in this work. For brevity, we label selected factors as highlighted in \textit{italics}. }
    \label{tab:factors_levels}
    \centering
    \scalebox{0.9}{
    \begin{tabular}{r|r|l}
        \hline
        Perspective & Factor & Levels \\
        \hline
        Temporal &
        Seasonality & 
            - \text{7-day \textit{MA}}\\
        &&  - \text{STL (\textit{trend})}\\
        \cline{2-3}
        &
        Granularity &
            - \text{50-day period}\\
        &&  - \text{10-day periods}\\
        \cline{2-3}
        & 
        Aggregation &
            - \textit{mean}\\
        &&  - \text{area under the curve (\textit{AUC})}\\
        &&  - \text{rank sums (\textit{RS})}\\
        \hline
        Multi-criteria &
        Comparison & 
            - \text{mean scalarized}\\
        &&  - \text{Pareto-dominance}\\
        &&  - \text{multiplicative $\epsilon$-dominance}\\
        \cline{2-3}
        &
        Tolerance ($\epsilon$) & $\{0.01, 0.05, 0.1\}$\\
        \hline
    \end{tabular}
    }
\end{table}

\paragraph{Seasonality effects} 
Temporal raw data is embedded with different sorts of seasonality. A straightforward approach to reduce such effects is to adopt moving averages~(MA). For instance, a \textbf{seven-day MA} spans over every day of the week, thus reducing weekday seasonality. However, the more days an MA comprises, the more it will propagate noise. As previously discussed, the relevance of noise in CMR data is such that Google recommends preprocessing them. In contrast to MA, more elaborate approaches such as seasonal-trend decomposition by loess~(\textbf{STL},~\cite{Cleveland1990stl}) attempt to isolate the \textbf{trend} of the data from both seasonal effects and noise. 

\paragraph{Discretization granularity}
Temporal discretization often leads to information loss regarding temporal dynamics. This loss is directly regulated by the temporal discretization granularity adopted, as a more fine-grained discretization still preserves some of these dynamics (at the cost of producing yet other time series). In this work, we analyze the data using both (i)~\textbf{a single time period}, comprising 50 days, and (ii)
~\textbf{multiple time periods}, comprising ten consecutive days each. In more detail, we have observed that for all localities considered, the period between the first restriction measure and the date when it was first relaxed comprises at least 50 days (but not much more than that). For this reason, after we center the data and apply seasonality approaches,%
\footnote{In a preliminary assessment, we have also considered restricting the time period of our analysis prior to applying seasonality approaches. However, we have observed that the outcomes of the techniques reflect better the original data when the whole series are considered.} 
we restrict our analysis to the first 50 days since the first restriction measure at each locality.%
\footnote{Considering days since a given event is a common option in COVID-19 analysis, and renders raw data further unsuited to the assessment. For instance, the 50-day period for two localities may differ as to the number of weekends they comprise, a factor that is critical in the context of mobility analysis.} 
If a coarse discretization is adopted, all 50 days comprise a single period. Instead, if a more fine-grained discretization is chosen, the analysis is performed on five periods comprising ten consecutive days each, producing a five-period time series for each aggregation technique we next discuss.

\paragraph{Temporal aggregation}
A straightforward approach to aggregate a time series is to compute its \textbf{mean}. A more ellaborate approach is to compute the area under the curve~(\textbf{AUC}) for each series, provided a common reference point.%
\footnote{The programming library we adopt in this work~(SciPy, \url{http://scipy.org}) does not allow us to specify a reference point for AUC computation. For this reason, before AUC aggregation we globally shift the data so it becomes non-negative and the origin of the y-axis can be used as reference point. Since CMR data is updated weekly and this could interfere with global shifting, we have performed shifting on data up to July 22nd, 2020.
}
Both mean and AUC can be regarded parametric approaches, as for a given day and category they are able to capture both (i)~which localities had larger mobility reduction and (ii)~the mobility reduction differences themselves. Conversely, a non-parametric approach such as rank sums~(\textbf{RS}) is only able to grasp the former. Effectively, an RS helps identify which localities were more consistently effective in mobility reduction than others, rather than by how much. 


\subsection{Multi-criteria decision making}

To assess CMR data from a multi-criteria perspective, we initally discuss how we prepare the data, and next the metrics we can employ to compare data from pairs or sets of localities.


\paragraph{Data preparation} 
Some of the most relevant recommendations from Google concerning CMR regard the particularities of the different places categories. In this work, we restrict our analysis to mobility in non-residential areas. Our rationale is that the indication that residential mobility should be maximized is not as clear as that mobility in non-residential places should be minimized. More precisely, Google does not specify the space granularity it adopts for residential places. As such, cultural traits such as social gatherings from neighbors that live in a same building would have to be minimized, and we do not know how the spatial discretization adopted by Google handles these scenarios. To ensure comparability for all the remaining place categories, prior to temporal aggregation we have scaled the data so that the variance from each category during the 50-day period considered becomes unitary, whatever the seasonality approach. Our motivation is two-fold. First, scaling the data from the 50-day period rather than the whole series ensures comparability between reduction rates. Second, scaling to unit variance rather than to fit a range as traditional in MCDM literature preserves the interpretability of the $\epsilon$-dominance analysis.

\begin{table}[!t]
\caption{Pareto dominance relations among localities~\cite{Zitzler2003performance}.}
\label{tab:dominance}
\centering
\scalebox{0.8}{
\begin{tabular}{p{.22\linewidth}cp{.72\linewidth}}
 \hline
  \textbf{Relation} &  \textbf{Notation} &  \textbf{Interpretation}\\
 \hline
 Strict\linebreak dominance	&   $l_{1} \prec\prec l_{2}$ 	& 	Mobility reduction at locality $l_{1}$ was larger than at locality $l_{2}$ for all place categories\\
 \hline
 Dominance		&	$l_{1} \prec l_{2}$		& 	Mobility reduction at locality $l_{1}$ was not smaller than at locality $l_{2}$ for all place categories, and was larger for at least one\\
\hline
 Weak\linebreak dominance &	$l_{1} \preceq l_{2}$		& 	Mobility reduction at locality $l_{1}$ was not smaller than at locality $l_{2}$ for all categories\\
\hline
 Indifference		&	$l_{1} \sim l_{2}$		& 	Locality $l_{1}$ presents the same mobility reduction as locality $l_{2}$ for all place categories\\
 \hline
 Incomparability	&	$l_{1} \mid\mid l_{2}$		& 	Neither $l_{1} \preceq l_{2}$ nor $l_{2} \preceq l_{1}$\\
 \hline
\end{tabular}
}
\end{table}

\paragraph{Multi-criteria comparison}
The most straightforward approach to comparing mobility from multiple localities is to aggregate the data from the different place categories, typically through an \textbf{scalarization} function. In this work, the \textbf{mean} of the multiple categories is an applicable scalarization function, since we do not pre-establish a preference over the place categories. However, aggregating the data from the different place categories incurs in information loss. The most commonly adopted alternatives to scalarization functions are \textbf{Pareto}- or $\epsilon$-\textbf{dominance}
~\cite{Zitzler2003performance,Laumanns2002epsilon}. The former is summarized in Table~\ref{tab:dominance}, and can discriminate situations where the mobility reduction from two localities are (i)~\textit{incomparable}, since each locality presents more effective mobility reduction for at least one of the place categories being assessed, and; (ii)~\textit{comparable}, when the mobility reduction for one locality is at least as effective as the mobility reduction for the other locality, for all place categories assessed. Complementarily, $\epsilon$-dominance is a relaxed form of Pareto dominance, where the decision maker can configure the tolerance level~($\epsilon$) of the analysis.

\paragraph{Dominance-compliant ranking}
When assessing a set of localities, it is interesting to rank them in a Pareto-compliant way. Among the best-established ranking approaches in MCDM theory is nondominated sorting~\cite{Goldberg1989genetic}, which produces the \textbf{dominance depth} measure~\cite{Bezerra2016}. The most relevant properties of dominance depth are that (i)~localities in a same cluster (same dominance depth) are incomparable and (ii)~localities in a cluster with larger dominance depth are dominated by some locality in a cluster with lower dominance depth. In this work, we evaluate localities based on their dominance depth, computed using the three different multi-criteria comparison approaches discussed above.%
\footnote{Note that dominance depth coupled with mean scalarization will most often produce singleton clusters, and is equivalent to ranking the mean values from the different place categories. Still, we model all comparison metrics using a common MCDM framework to have the ranking based on mean scalarization as baseline.}


\medskip
\section{Results}
\label{sec:results}

To empirically demonstrate the methodology discussed in the previous section, we conduct both a region- and a country-level mobility reduction analysis for selected localities. Concretely, we start with an analysis of regions that experienced a severe COVID-19 outbreak, and for simplicity we limit this analysis to the two most recent global epicentres: (i)~Europe, from which we include Lombardia~(Italy) and Île-de-France~(France), and; (ii)~the Americas, from which we consider the states of New York~(USA) and Amazonas and São Paulo~(Brazil).%
\footnote{Differently from other countries, in Brazil the epicenter~(São Paulo) was not the region with the most severe outbreak~(Amazonas)~\cite{jhu}.} 
In common, mobility restriction measures adopted by those countries were originally proposed on a regional basis. By contrast, other countries enforced nation-wide restrictions early on, which we assess with the country-level analysis. Besides Europe~(Germany and Spain) and the Americas~(Argentina and Canada), the country-level analysis also includes Asia~(Japan and South Korea) and Oceania~(New Zealand). 

Next, we first illustrate the effects of the first steps of the methodology discussed in Section~\ref{sec:methodology}, specifically data calibration and seasonality approaches. Later, we present our region- and country-level analyses.


\subsection{Preliminary discussion}
\begin{figure}[!t]
    \centering
	\includegraphics[width=\linewidth]{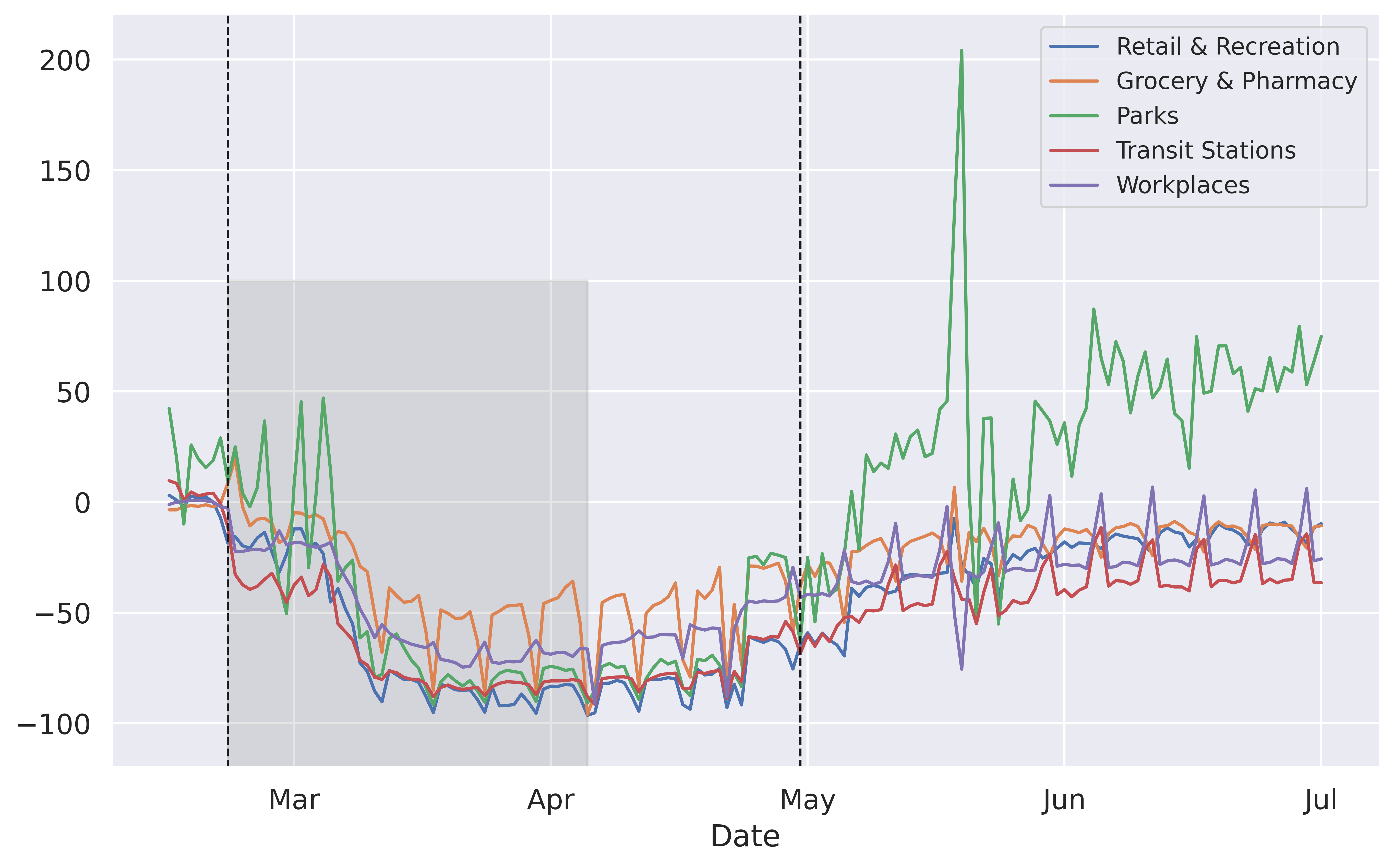}
	\caption{Lombardia's mobility data~(y-axis) for all categories before data calibration. Left-most vertical dashed line: first mobility restriction~(Feb~23rd). Right-most vertical dashed line: restrictions were first relaxed~(May~11th). Shaded region: Lombardia's 50-day period considered in the assessment.}
	\label{FIG:region_lombardy}
\end{figure}

As discussed in the CMR guidelines~\cite{CMR} and also in Section~\ref{sec:methodology}, data from different localities and/or categories may need to be calibrated. This is illustrated in Figure~\ref{FIG:region_lombardy}, where we depict mobility data for the different place categories in Lombardia. In the plot, the left-most vertical dashed line indicates the date of the first mobility restriction adopted in Lombardia~(February 23rd), whereas the right-most vertical dashed line indicates the date when restrictions were first relaxed~(May 11th). The shaded region highlights Lombardia's 50-day period considered in this assessment. For the dates before the first restriction, the mobility at the different categories is in general centered around zero, indicating that the baseline period adopted by Google (January 3rd to February 6th)
can be regarded representative of an average mobility scenario in Lombardia. However, this is not the case for the \textit{Parks} category, for which the mobility before the first restriction measure is higher than on the baseline period. 

\begin{figure}[!t]
    \centering
	\includegraphics[width=\linewidth, clip=true, trim=30px 0 0 0]{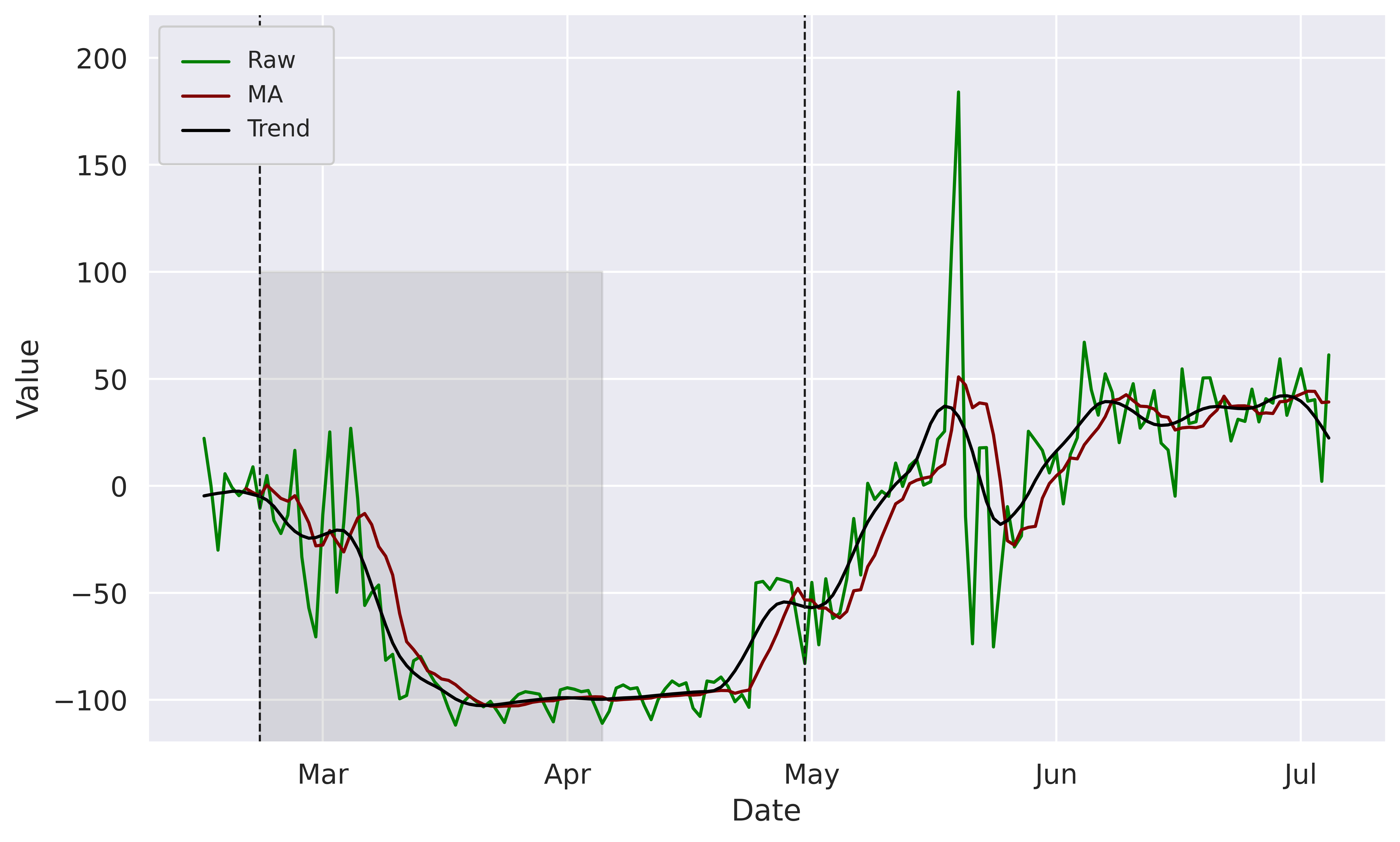}
	\caption{Mobility data from \textit{Parks} in Lombardia after calibration, comparing different seasonality approaches (MA and \textit{trend}) with the original raw data.}
	\label{FIG:see-lombardia-parks}
\end{figure}

\newcommand{\blueten}{\cellcolor{blue!10}}
\newcommand{\bluetwenty}{\cellcolor{blue!20}}
\newcommand{\bluethirty}{\cellcolor{blue!30}}
\newcommand{\bluethirtyfive}{\cellcolor{blue!35}}
\newcommand{\blueforty}{\cellcolor{blue!40}}
\newcommand{\bluefifty}{\cellcolor{blue!55}}

\newcommand\colorgd[1]{\cellcolor{blue!#10}#1}

\begin{table*}[!t]
\begin{minipage}{0.25\textwidth}
    \centering
    \scalebox{0.8}
    {
    \begin{tabular}{c|c|c|}
        \cline{2-3}
        & \multicolumn{2}{c|}{AUC/Mean/RS} \\ \cline{2-3} 
        & MA & Trend  \\ \hline
        \multicolumn{1}{|c|}{Île-de-France} & \colorgd{1} & \colorgd{1}  \\ \hline
        \multicolumn{1}{|c|}{Lombardia} & \colorgd{2} & \colorgd{2} \\ \hline
        \multicolumn{1}{|c|}{Amazonas} & \colorgd{5} & \colorgd{4}  \\ \hline
        \multicolumn{1}{|c|}{São Paulo} & \colorgd{3}  & \colorgd{3}  \\ \hline
        \multicolumn{1}{|c|}{New York} & \colorgd{4} & \bluefifty5  \\ \hline
    \end{tabular}%
    }
    \smallskip
    \footnotesize{\\(a) \textit{Workplaces}}
    \label{tab:workplaces}
\end{minipage}
\begin{minipage}{0.30\textwidth}
\centering
\scalebox{0.8}
{
\begin{tabular}{c|c|c||c|c|}
\cline{2-5}
& \multicolumn{2}{c||}{AUC/Mean} &  \multicolumn{2}{c|}{RS} \\ \cline{2-5} 
  & MA    & Trend      & MA    & Trend       \\ \hline
\multicolumn{1}{|c|}{Île-de-France} & \colorgd{1} & \colorgd{2} & \colorgd{1} & \colorgd{1}  \\ \hline
\multicolumn{1}{|c|}{Lombardia} & \colorgd{2} & \colorgd{1} & \colorgd{2} &\colorgd{2} \\ \hline
\multicolumn{1}{|c|}{Amazonas} & \colorgd{3}  & \colorgd{3} & \colorgd{3} & \colorgd{3}  \\ \hline
\multicolumn{1}{|c|}{São Paulo} & \colorgd{4} & \colorgd{4} & \colorgd{4} & \colorgd{3}  \\ \hline
\multicolumn{1}{|c|}{New York} & \colorgd{5} & \colorgd{5} & \colorgd{5} & \colorgd{5}  \\ \hline
    \end{tabular}%
    }
    \smallskip
    \footnotesize{\\(b) \textit{Grocery \& Pharmacy}}
\label{tab:grocery_parmacy}
\end{minipage}
\begin{minipage}{0.45\textwidth}
\centering
\scalebox{0.8}
{
\begin{tabular}{c|c|c||c|c||c|c|}
\cline{2-7}
& \multicolumn{2}{c||}{AUC} & \multicolumn{2}{c||}{Mean} & \multicolumn{2}{c|}{RS} \\ \cline{2-7} 
   & MA    & Trend      & MA    & Trend       & MA    & Trend    \\ \hline
\multicolumn{1}{|c|}{Île-de-France} & \colorgd{1} & \colorgd{2} & \colorgd{2} & \colorgd{2} & \colorgd{1} & \colorgd{1}   \\ \hline
\multicolumn{1}{|c|}{Lombardia} & \colorgd{2} & \colorgd{1} & \colorgd{1} &\colorgd{1} & \colorgd{2} & \colorgd{2} \\ \hline
\multicolumn{1}{|c|}{Amazonas} & \colorgd{3}  & \colorgd{3} & \colorgd{3} & \colorgd{3} & \colorgd{3} & \colorgd{3}  \\ \hline
\multicolumn{1}{|c|}{São Paulo} & \colorgd{4} & \colorgd{4} & \colorgd{4} & \colorgd{4} & \colorgd{4} & \colorgd{4}  \\ \hline
\multicolumn{1}{|c|}{New York} & \colorgd{5} & \colorgd{5} & \colorgd{5} & \colorgd{5} & \colorgd{5} & \colorgd{5}  \\ \hline
    \end{tabular}%
    }
    \smallskip
    \footnotesize{\\(c) \textit{Parks}}    
\end{minipage}
    \caption{Regions respectively ranked according to their mobility in (a)~\textit{Workplaces}, (b)~\textit{Grocery \& Pharmacy}, and (c)~\textit{Parks}, aggregated over a 50-day period using different measures (AUC, mean and RS) and seasonality approaches~(MA and \textit{trend} data). Regions are sorted to match the ranking most frequently observed.}
\label{tab:parks}
\end{table*}


By contrast, Figure~\ref{FIG:see-lombardia-parks} depicts the mobility data from \textit{Parks} in Lombardia after data is calibrated~(green line), and also illustrates the effect of seasonality approaches applied post-calibration. Specifically, data processed using seven-day moving averages is referred to as MA~(brown line), and the trend component extracted with the STL approach is referred to as \textit{trend}~(black line). Concerning calibration, the mobility data from \textit{Parks} in Lombardia before the first mobility restriction now presents zero mean. As a consequence, we observe that the whole series has been shifted down along the y-axis, and the mobility reduction from \textit{Parks} during the 50-day period assessed is now often greater than from the remaining place categories. Regarding seasonality approaches, data processed using moving averages is less smooth than data processed with the STL approach. As discussed in Section~\ref{sec:methodology}, STL is a technique able to partially mitigate the presence of noise, whereas moving averages may propagate it. Indeed, in comparison to the remaining place categories given in~Figure~\ref{FIG:region_lombardy}, we notice that \textit{Parks} is a category where mobility follows a less-structured pattern, being more prone to exceptional circumstances~(noise). 

\subsection{Region-level analysis}

After calibration and seasonality processing, we proceed to the analysis of the selected regions. For clarity, we group the main insights observed following the remainining steps of the methodology discussed in Section~\ref{sec:methodology}.

\subsubsection{Temporal aggregation}
We begin our discussion with the temporal aggregation of the 50-day series for each region and place category. The three different aggregation approaches~(AUC, mean, and RS) were computed using MA and \textit{trend} data. Table~\ref{tab:parks} shows regions ranked for each of these combinations when considering mobility data from \textit{Workplaces}~(a), \textit{Grocery \& Pharmacy}~(b), and \textit{Parks}~(c). For clarity, regions are sorted to match the ranking most frequently observed in the table, namely: Île-de-France, Lombardia, Amazonas, São Paulo, and New York. For brevity, results for categories \textit{Transit Stations} and \textit{Retail \& Recreation} are provided as supplementary material~\cite{CavOliWanBez2020ss}, as the mobility data from these categories are highly correlated to mobility data from \textit{Workplaces}~\cite{ritchie2020}.

\begin{figure}[!t]
    \centering
	\includegraphics[width=\linewidth]{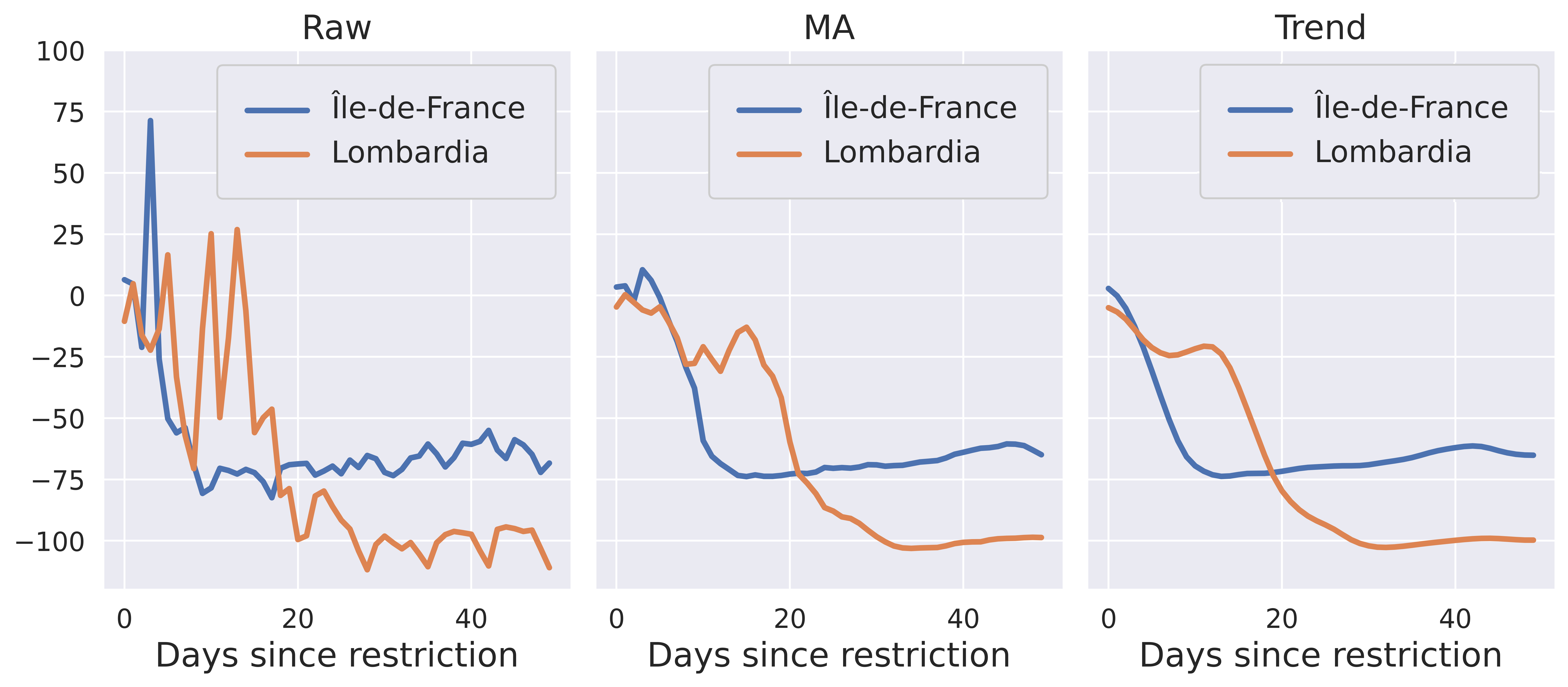}
	\caption{Mobility data from \textit{Parks} comparing Île-de-France and Lombardia after calibration, with different seasonality approaches, zoomed at the 50-day period considered in the assessment.}
	\label{FIG:seasonality-lombardia-idf-parks}
\end{figure}

Regardless of the aggregation measure or seasonality approach adopted, we can see two distinct clusters on Table~\ref{tab:parks}. The first cluster comprises Île-de-France and Lombardia, which alternate between the first and second positions. The second cluster comprises Amazonas, São Paulo, and New York, which alternate among the third, fourth, and fifth positions. Differences in rankings are least frequent for \textit{Workplaces}, for which aggregation measures agree, and most frequent for \textit{Parks}, for which both aggregation measures and seasonality approaches interact.
To illustrate these effects, Figure~\ref{FIG:seasonality-lombardia-idf-parks} shows the 50-day series from \textit{Parks} in Lombardia~(orange line) and Île-de-France~(blue line) with different seasonality approaches. As discussed in the previous section, MA and \textit{trend} present different smoothness degrees due to how noise is handled. The different shapes of the curves affect AUC aggregation, and results for MA diverge from results with \textit{trend} data. Regarding mean aggregation, rankings for the different seasonality approaches match in the \textit{Parks} category, though they diverge for \textit{Grocery \& Pharmacy}. This is also the case for RS aggregation, which Fig.~\ref{FIG:seasonality-lombardia-idf-parks} also helps illustrate. In particular, the shape differences seen for \textit{Parks} do not strongly alter the relative rankings of the regions, and so RS aggregation is not affected. We conclude from these results that AUC aggregation is more sensitive to the choice of seasonality approach than the remaining aggregation measures considered.

\subsubsection{Discretization granularity}
\begin{figure}[!t] 
    \centering
	\includegraphics[scale=.25, clip=true, trim=0 90px 0  0]{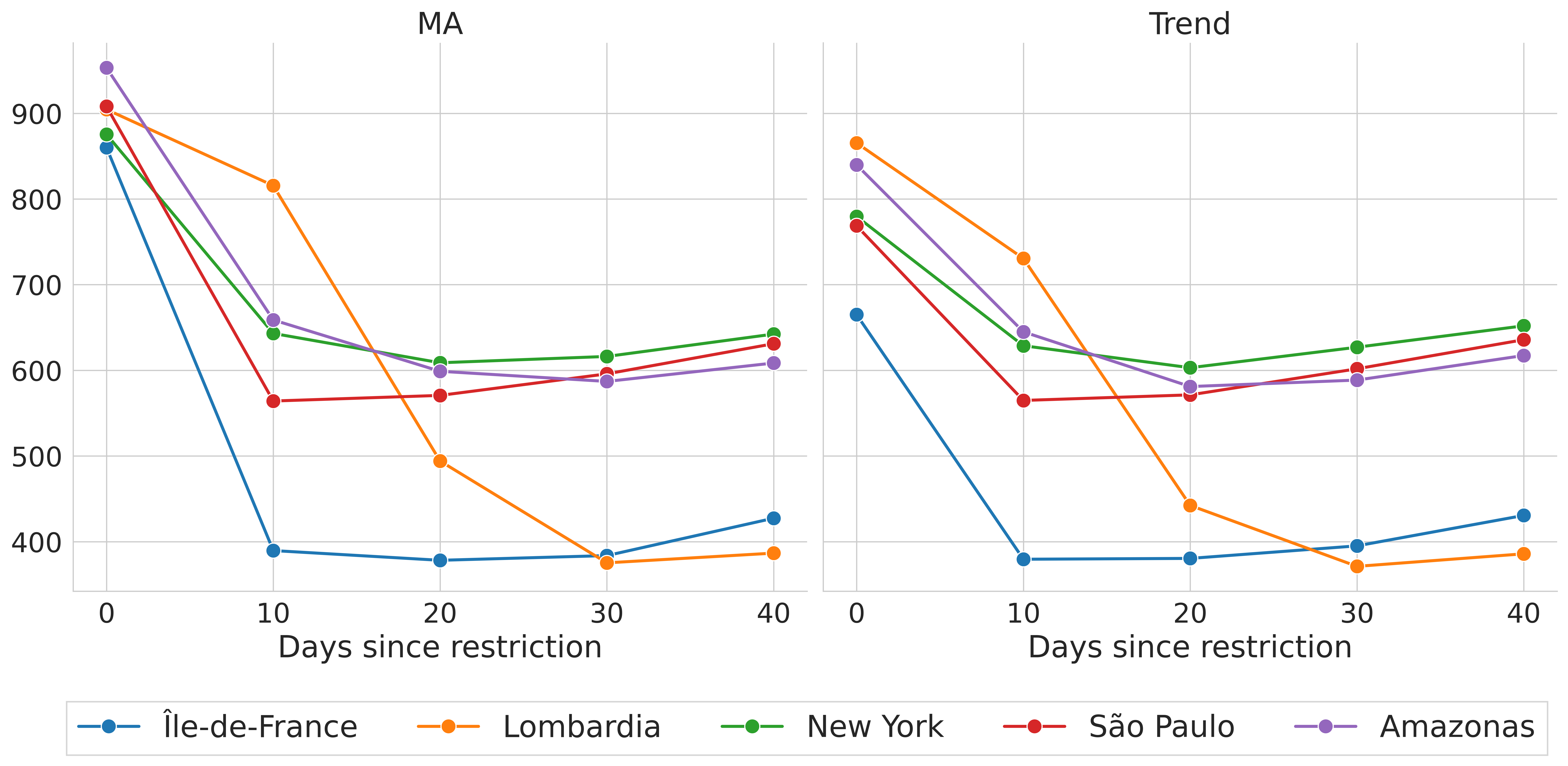}
	\label{FIG:regions-long_results_complete_10D_auc}
    \centering
	\includegraphics[scale=.25, clip=true, trim=0 0 0 30px]{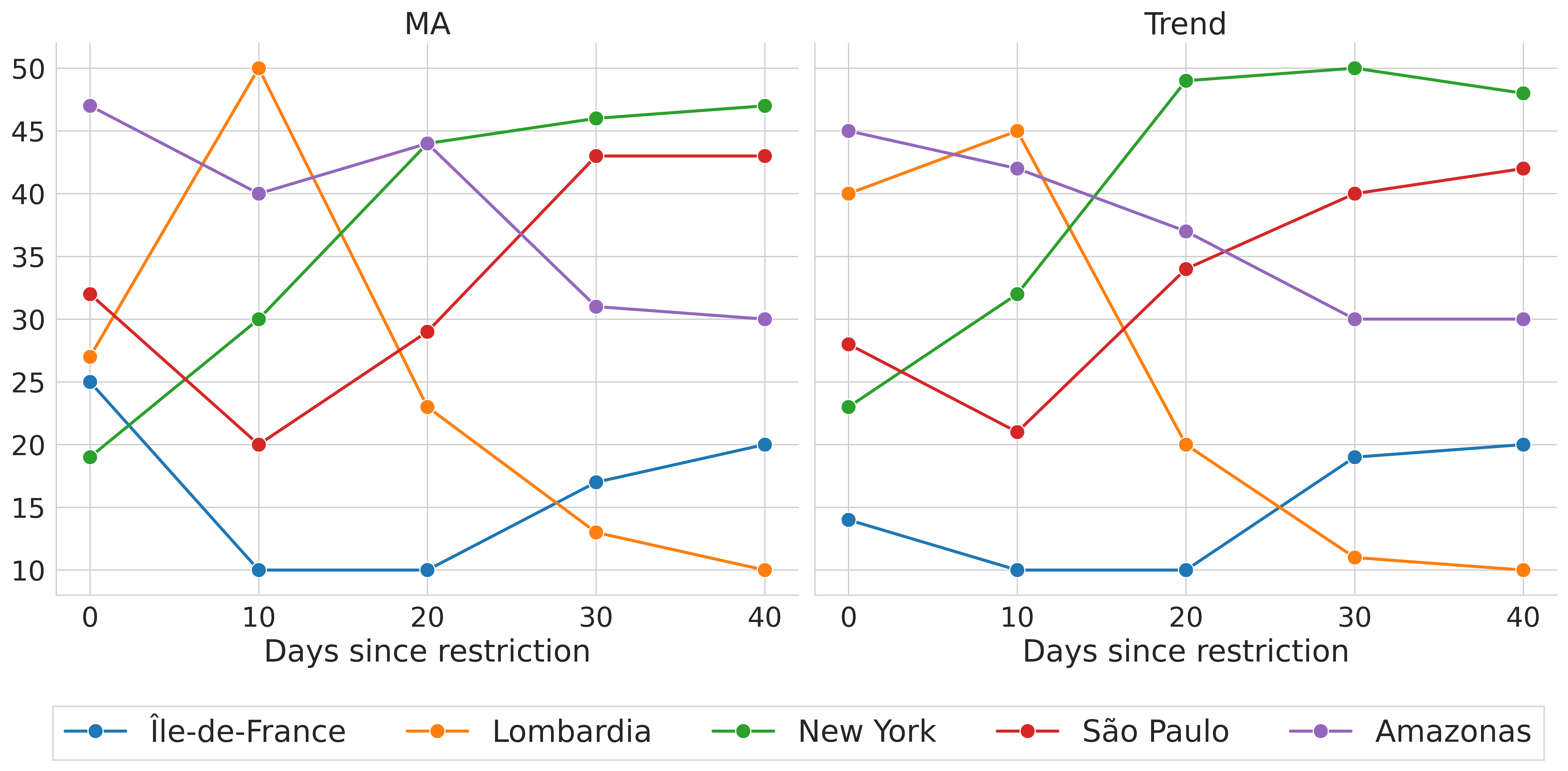}
	\caption{Multiple-time-period analysis for \textit{Workplaces} using MA and \textit{trend} data and different aggregation measures. Top: AUC. Bottom: RS. Each time period is represented by its initial day, given as a marker.}
	\label{FIG:regions-long_results_complete_10D_rank}
\end{figure}

Time series assessment using different levels of temporal discretization granularity often provide complementary insights, at the cost of producing novel time series. To assess the influence of time granularity, in this section we compute aggregated measures for five time periods, comprising ten consecutive days each. Concretely, given an individual place category and region, we compute aggregated measures for each time period considering MA and \textit{trend} data. For brevity, we illustrate the resulting time series for category \textit{Workplaces} in Figure~\ref{FIG:regions-long_results_complete_10D_rank}, which depicts series produced using AUC~(top) and RS~(bottom) aggregation. In particular, we omit the series produced by mean aggregation as they are roughly equivalent to the series produced using AUC aggregation for this category. For each plot, the x-axis represents the 50-day period divided into 10-consecutive-day periods, labeled according to the number of days since the first restriction measure in the given region. 


Figure~\ref{FIG:regions-long_results_complete_10D_rank} reveals interesting insights about the dynamics of mobility reduction in Lombardia and Amazonas~(purple line). In more detail, the single-time-period analysis given in Table~\ref{tab:parks}a shows Lombardia ranking worse only than Île-de-France. Conversely, when multiple time periods are considered, the reduction observed in Lombardia during the first and second time period were among the worst of the regions considered. As of the third time period on, we observe a significant improvement in mobility reduction in Lombardia, eventually leading to a better rank than all the remaining regions for all combinations of aggregation and seasonality approaches. A similar pattern is observed for Amazonas, which often was ranked worse than all regions but New York~(green line) in Table~\ref{tab:parks}a. When multiple time periods are considered, we see that Amazonas actually starts worse than all other regions~(first time period), but improves its mobility reduction rate to the point of even surpassing the reduction seen in São Paulo~(red line). 



\subsubsection{MCDM analysis}

To compare mobility from different localities taking into account all place categories simultaneously, we employ the different multi-criteria comparison approaches discussed in Section~\ref{sec:methodology}: mean scalarization and Pareto- and $\epsilon$-dominance. Furthermore, to produce a ranking of the regions, we couple these different comparison approaches with nondominated sorting, i.e., the ranking produced depicts the dominance depth of each region. For simplicity, we start this analysis considering a single 50-day time period, and group the most relevant insights we discuss to match the structure used in Section~\ref{sec:methodology}.

\begin{figure}[!t]
    \centering
	\includegraphics[width=.90\linewidth,clip=true, trim=0 0px 0 0]{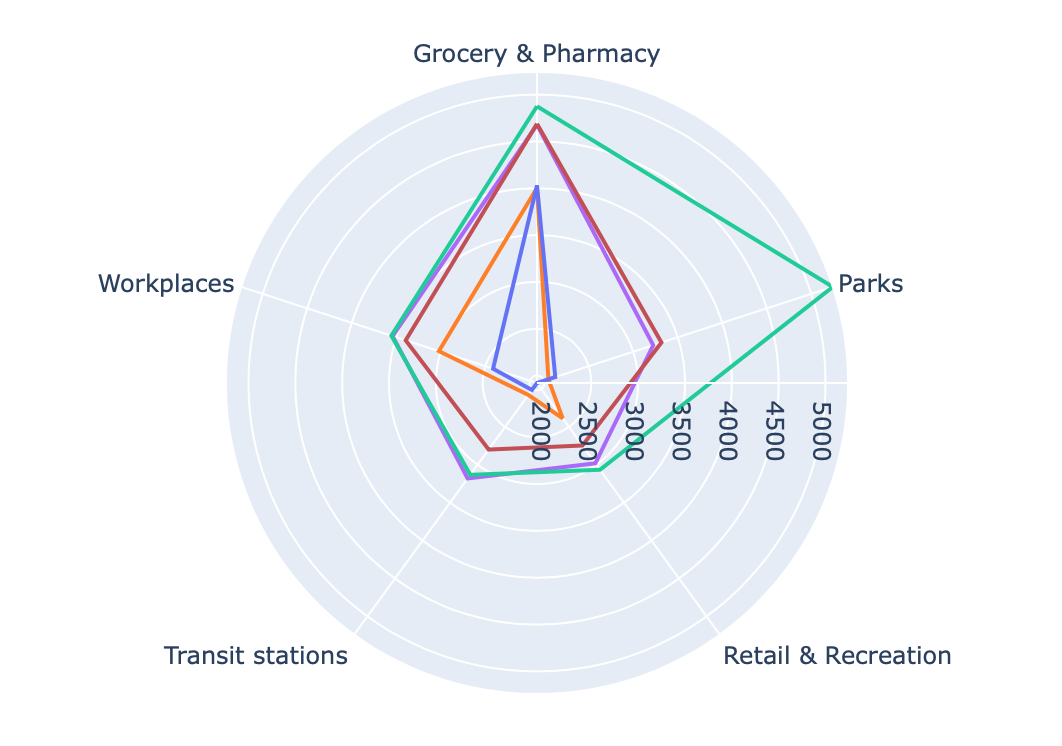}
	\includegraphics[width=.95\linewidth]{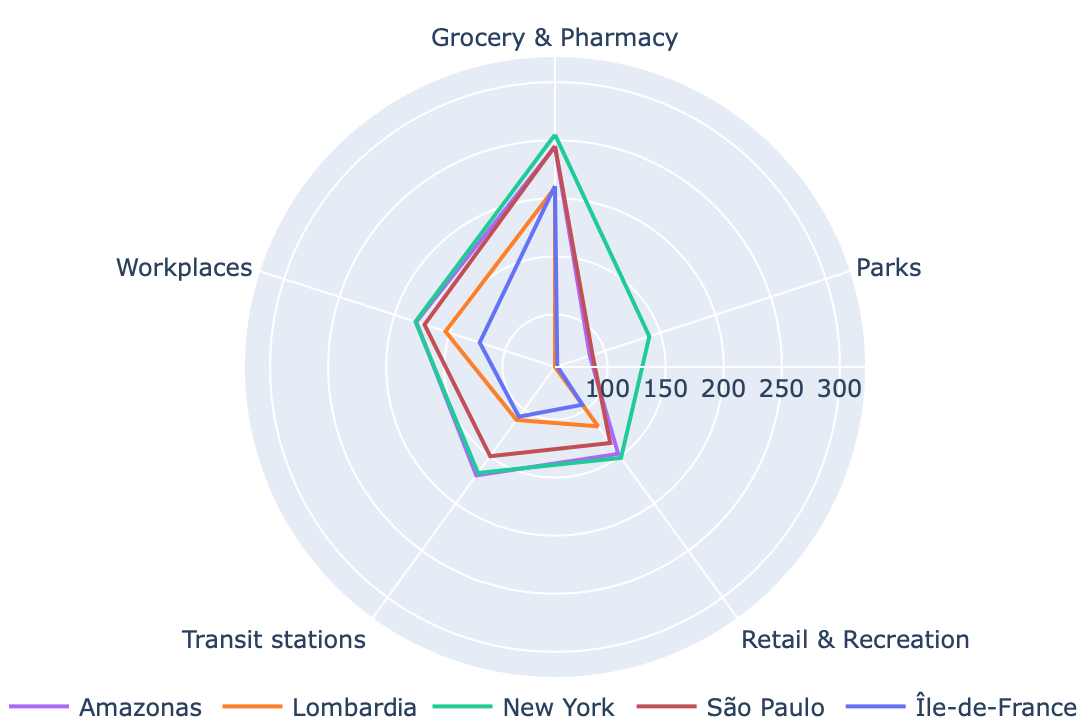}
	\caption{Multi-criteria visualization of AUC-aggregated mobility \textit{trend} data comparing different regions and categories. Top: original. Bottom: scaled.}
	\label{FIG:regions_scaling}
\end{figure}

\paragraph{Data preparation}
Scaling effects are illustrated in Fig.~\ref{FIG:regions_scaling}, where a radar chart depicts AUC aggregation of the original~(top) and scaled~(bottom) \textit{trend} data. In more detail, this polar coordinate plot depicts (i)~categories as angles and (ii)~mobility reduction aggregated over the 50-day period as \textit{radii}. More importantly, this multi-dimensional visualization simplifies the multi-criteria comparison between different localities, as locality $l_1$ dominates locality $l_2$ iff the polygon given for $l_1$ is contained by the polygon given for $l_2$. Besides the improved balance between categories, we observe that scaling may also affect dominance depending on the seasonality approach and/or aggregation measure considered. For instance, Fig.~\ref{FIG:regions_scaling}~(top) shows that only mobility reduction rates from categories \textit{Grocery \& Pharmacy} and \textit{Parks} were deemed greater in Lombardia than in Île-de-France, rendering both localities Pareto-incomparable. Yet, scaling employed prior to aggregation~(Fig.~\ref{FIG:regions_scaling}, bottom) renders reduction rates from all categories greater in Île-de-France than in Lombardia, and so the former dominates the latter.

\paragraph{Multi-criteria comparison}
Figure~\ref{FIG:regions_scaling}~(bottom) also illustrates the different multi-criteria comparison approaches we consider. For instance, AUC-aggregated mobility reduction rates in New York are (marginally) larger than in Amazonas for categories \textit{Retail \& Recreation}, \textit{Transit stations}, and \textit{Workplaces}. According to Pareto-dominance, these two localities are incomparable. Conversely, the difference from these categories is so small that $\epsilon$-dominance considers that Amazonas dominates New York for all tolerance levels investigated in this work.

\paragraph{Dominance-compliant ranking}
Table~\ref{tab:dominance_all_categories_df} gives rankings computed using Pareto-dominance. In general, Lombardia and Île-de-France comprise the cluster with largest mobility reduction rates. By contrast, Amazonas and São Paulo are now a cluster that always dominates New York. Overall, two rankings can be observed, differing as to whether Île-de-France dominates Lombardia or not. As expected, mean  and RS aggregation are not affected by seasonality approach for this matter, whereas AUC is.
In contrast to dominance-based analysis, ranking regions using mean scalarization produces the same total order, independently of the aggregation measures and seasonality approaches applied: Île-de-France, Lombardia, São Paulo, Amazonas, and New York. This order differs from the most frequent in Table~\ref{tab:parks} in that São Paulo precedes Amazonas, but matches the order seen for \textit{Workplaces} when trend data is adopted. Besides scaling, this could also be a consequence of the number of categories in CMR data correlated with \textit{Workplaces}.

\begin{table}[!t]
\centering
\scalebox{0.8}{
\begin{tabular}{c|c|c||c|c||c|c|}
\cline{2-7}
& \multicolumn{2}{c||}{AUC} & \multicolumn{2}{c||}{Mean} & \multicolumn{2}{c|}{RS} \\ \cline{2-7}
& MA    & Trend   & MA    & Trend    & MA    & Trend    \\ \hline
\multicolumn{1}{|c|}{Lombardia} & \colorgd{1} & \colorgd{2}  & \colorgd{1} & \colorgd{1}  & \colorgd{2} & \colorgd{2} \\ \hline
\multicolumn{1}{|c|}{Île-de-France} & \colorgd{1} & \colorgd{1} & \colorgd{1} & \colorgd{1} &  \colorgd{1}  & \colorgd{1} \\ \hline
\multicolumn{1}{|c|}{Amazonas} & \colorgd{2} & \colorgd{3}  & \colorgd{2} & \colorgd{2}  & \colorgd{3} & \colorgd{3} \\ \hline
\multicolumn{1}{|c|}{São Paulo} & \colorgd{2} & \colorgd{3} & \colorgd{2} & \colorgd{2}  & \colorgd{3} & \colorgd{3} \\ \hline
\multicolumn{1}{|c|}{New York}  & \colorgd{3} & \colorgd{4} & \colorgd{3} & \colorgd{3}  &\colorgd{4} &\colorgd{4} \\ \hline
\end{tabular}%
}
\caption{Regions ranked according to Pareto-dominance.}
\label{tab:dominance_all_categories_df}
\end{table}

\paragraph{Tolerance configuration}
When tolerance is set very low~($\epsilon = 0.01$), the only difference to Pareto-dominance rankings concerns mean aggregation of MA data, for which Île-de-France now $\epsilon$-dominates Lombardia. Conversely, a tolerance level set too high~($\epsilon = 0.1$) renders AUC/mean rankings identical to mean-scalarized rankings. Concerning RS, São Paulo $\epsilon$-dominates Amazonas and New York, which are now considered $\epsilon$-incomparable.
The setup using $\epsilon = 0.05$ shares characteristics from both other setups. Specifically, AUC/mean rankings match mean-scalarized rankings, whereas RS rankings match Pareto-dominance rankings. We conclude from these results that RS aggregation is less sensitive to $\epsilon$-dominance.

\begin{figure}[!t]
\centering
	\includegraphics[width=\linewidth]{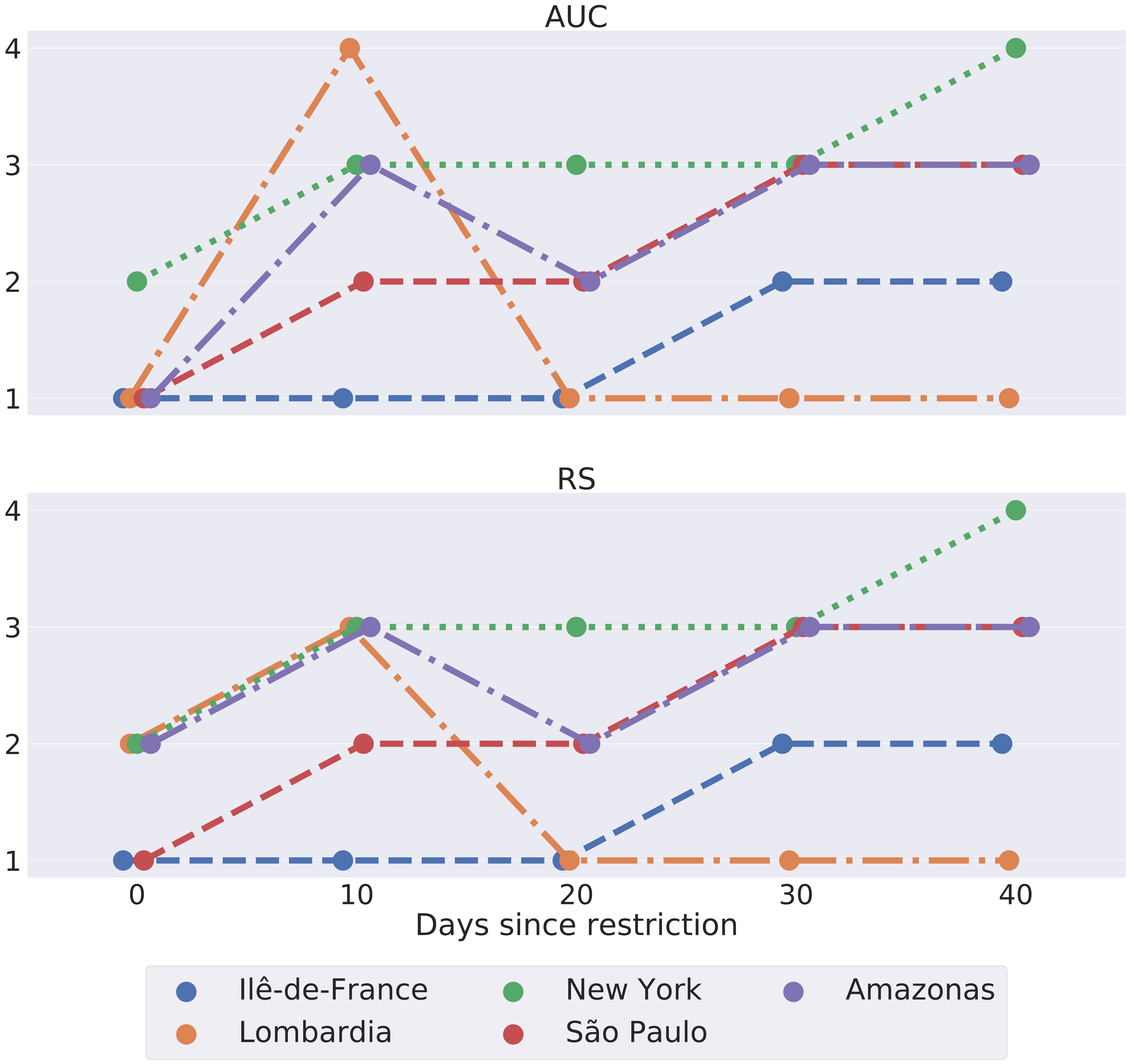}
	\caption{Parallel coordinates plot depicting the changes in Pareto-compliant rankings of the different regions. Each row represents a different aggregation measure applied to scaled trend data. Top: AUC. Bottom: RS.}
	\label{FIG:regions_results_complete_10D_depth}
\end{figure}

\medskip
Finally, we assess insights from the multiple-time-period analysis, in which we investigate the effect of the temporal discretization granularity.  Figure~\ref{FIG:regions_results_complete_10D_depth} depicts a parallel coordinates plot for this assessment. Each of the five time periods is represented as a vertical axis, and each region is represented by a polyline with vertices on the axes. The position of each vertex corresponds to the dominance depth of the given region in the corresponding time period. For clarity, vertices that would collide are slightly shifted along the x-axis. Furthermore, only the analysis for scaled \textit{trend} data is shown, as we next discuss.

\paragraph{Experimental factor effects}
We observe ranking variability as a function of the seasonality approach, aggregation measure, and time period considered. Specifically, we notice that rankings vary less as a function of experimental factors as time progresses. Indeed, for the first time period we observe four different rankings, whereas for the latter three a single ranking per period is produced by all aggregation measures and seasonality approaches. Overall, we recommend that multiple-period analyses consider only scaled \textit{trend} data as we do, given the benefits demonstrated by these factors.

\paragraph{Temporal dynamics}
Figure~\ref{FIG:regions_results_complete_10D_depth} depicts AUC and RS analyses of the scaled \textit{trend} data. In particular, mean aggregation rankings equal RS rankings, and for brevity they are not shown. As previously discussed, disagreements in rankings are observed only on the two initial time periods. Interestingly, Île-de-France and São Paulo present low dominance depths on those periods regardless of the aggregation measure considered. This contrasts with the later time-period rankings (and also single-period rankings), in which São Paulo is generally part of a cluster with higher dominance depth, being often incomparable w.r.t. Amazonas and/or New York. Conversely, Lombardia is often ranked poorly in the early time-period rankings, presenting lower dominance depth only from the third period on. Altogether, these observations confirm the importance of bridging sound TSA and MCDM methodology.

\subsection{Country-level results}

In the previous section, we have discussed how the different methodologies considered have affected results. In this section, we focus on an MCDM analysis of the different countries selected, highlighting whether insights discussed in the region-level analysis are still observable. For simplicity, we start our analysis considering a single time period comprising 50 days.

\begin{figure}[!t]
    \centering
    \includegraphics[width=0.83\linewidth, clip=true, trim=0 10px 0 15px]{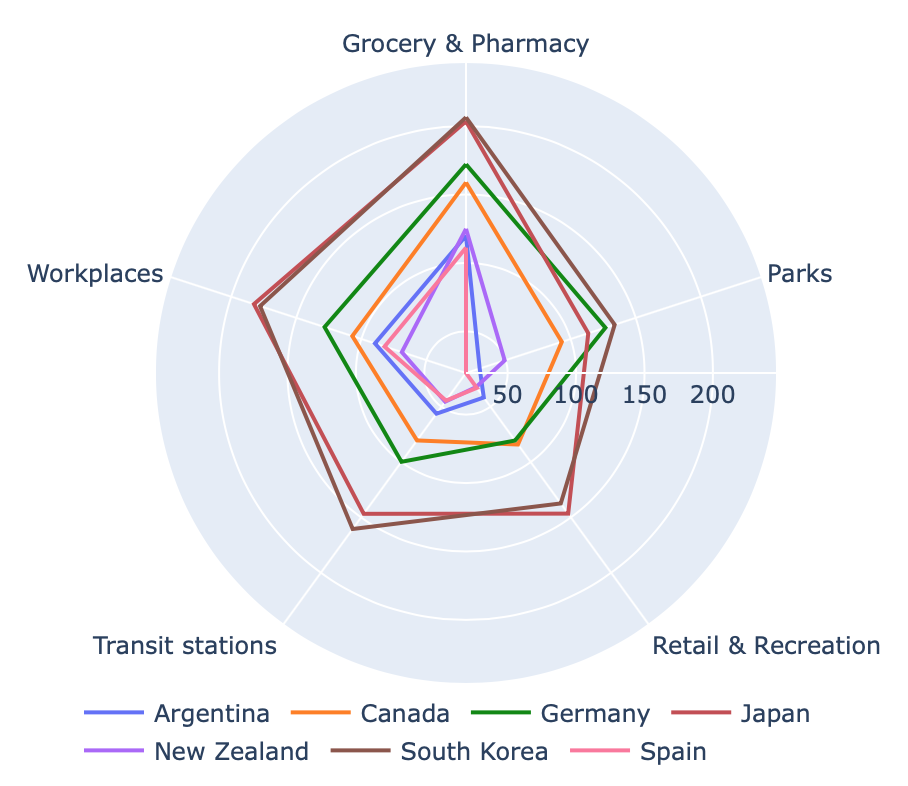}
	\caption{Multi-criteria visualization of AUC-aggregated scaled \textit{trend} data comparing different countries and categories.}
	\label{FIG:countries_scaling}
\end{figure}

\paragraph{Scaling and comparison}
Figure~\ref{FIG:countries_scaling} shows a radar chart for the AUC-aggregated mobility using scaled \textit{trend} data, which is representative of results seen for other factors. Once again, scaling balances the importance of the categories, but this time it does not interfere with the multi-criteria comparison. At a first glance, we observe in Fig.~\ref{FIG:countries_scaling}  two groups of disjoint countries, namely (i)~Spain, New Zealand, and Argentina, and (ii)~Canada, Germany, Japan, and South Korea. Within each group, we notice that some differences are relatively small, which impacts the later discussion on $\epsilon$-dominance.

\paragraph{Dominance-compliant ranking}
Both Pareto- and \text{$\epsilon$-dominance} with $\epsilon = 0.01$ converge to a same ranking between countries, whatever the factors considered. The cluster with lowest dominance depth comprises Spain~(pink) and New Zealand~(purple), followed by a singleton comprising Argentina~(blue). Figure~\ref{FIG:countries_scaling} helps visualize why dominance depth indicates that these countries comprise two clusters. Specifically, even if New Zealand is incomparable w.r.t. to Spain and Argentina, Spain dominates Argentina. 
The last two clusters comprise (i)~Canada and Germany and (ii)~Japan and South Korea. Once again, a closer look at Fig.~\ref{FIG:countries_scaling} confirms that Canada~(orange) dominates Japan~(red) and South  Korea~(brown). This way, Germany~(green) is clustered with Canada, even if Germany and Japan are incomparable. Finally, these multi-criteria nuances would have been completely overlooked if mean scalarization had been employed, resulting on a total order of the countries: Spain, New Zealand, Argentina, Canada, Germany, Japan, and South Korea. 


\begin{figure}[!t]
\centering
	\includegraphics[width=\linewidth]{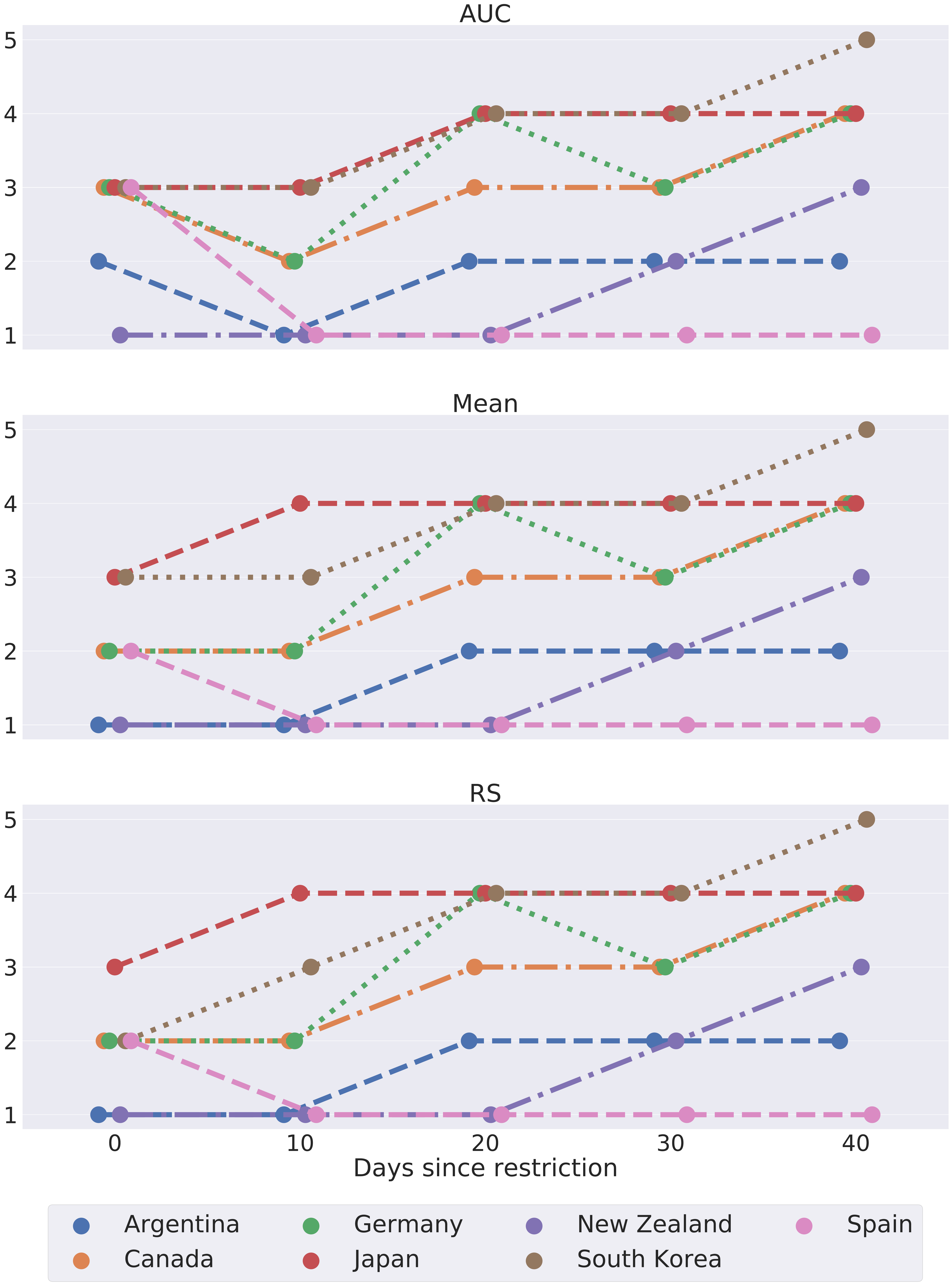}
	\caption{Parallel coordinates plot depicting the changes in Pareto-compliant rankings of the different countries over time. Each row represents a different aggregation measure applied to scaled \textit{trend} data.}
	\label{FIG:countries_results_complete_10D_depth}
\end{figure} 

\paragraph{Tolerance configuration}

As discussed above, a small tolerance value~($\epsilon = 0.01)$ is unable to affect rankings. The remaining insights observed match to some extent what had been discussed for the region-level analysis. In more detail, AUC and mean rankings for $\epsilon \in \{0.05,0.1\}$ show that Canada $\epsilon$-dominates Germany. As a result, Canada now comprises a singleton cluster, and two new clusters are formed: (i)~Germany and Japan, and (ii)~South Korea. This is also illustrated in Figure~\ref{FIG:countries_scaling}, where we can see that Japan is incomparable w.r.t. both Germany and South Korea, but the former dominates the latter. As in the region-level analysis, RS aggregation is less sensitive to $\epsilon$-dominance, with the only change in rankings observed concerning $\epsilon = 0.1$ and variying as a function of the seasonality approach adopted. Specifically, rankings computed using MA data remain the same, whereas rankings computed using \textit{trend} data indicate that Japan $\epsilon$-dominates South Korea.

\medskip
We conclude this analysis changing the temporal granularity adopted. As with the previously discussed multiple-time-period region-level analysis, experimental factors strongly affect rankings specially for the initial time periods. Figure~\ref{FIG:countries_results_complete_10D_depth} depicts the parallel coordinate plots using different aggregation measures of scaled \textit{trend} data. In common, rankings for all measures agree as of the third time period, and also on the second time period for mean~(middle) and RS~(bottom) aggregation. The most significant differences are observed for the first time period, on which none of the aggregation measures agree.

Concerning mobility dynamics, it is interesting to remark that Spain is never ranked in the lowest depth cluster on the first time period, whatever the aggregation measure considered. By contrast, New Zealand presented a strong mobility reduction rate in the early periods, but was later surpassed by Spain and even Argentina. Regarding the remaining countries, it is interesting to observe how South Korea starts in a same-~(mean) or in a lower-depth~(AUC and RS) cluster than Japan, but ends in a higher-depth cluster whichever measure considered. Finally, we remark that none of the rankings for a given time period match the Pareto-dominance rankings for the single-time-period analysis, which corroborates the importance of an assessment that considers multiple discretization granularities.

\section{Conclusion}
\label{sec:conclusion}

The novel coronavirus disease~(COVID-19) has been the most serious public health concern of the recent decades due to its quick spread across the world. In the absence of pharmaceutical alternatives, social distancing has been a key approach for countries trying to control the pandemic. Yet, social distancing is economically expensive, and countries have adhered to this measure in different rates.
The community mobility reports~(CMR) produced by Google weekly are a relevant example of how technology companies have assisted in the fight against COVID-19. Fully grasping the insights they provide requires a multi-disciplinary approach, comprising domains as diverse as time series analysis~(TSA), data visualization, and multi-criteria decision making~(MCDM). More importantly, the speed with which the scientific community is being pressed to produce results makes methodological soundness yet more important.

In this work, we have discussed MCDM concepts that enable a multi-criteria  analysis of CMR data. Specifically, we have discussed how to (i)~ensure comparability between the mobility reduction rates from different place categories, (ii)~compare the mobility reduction rates from a pair of localities through different Pareto dominance relations, and; (iii)~cluster and rank a set of localities based on their mobility reduction rates in a Pareto-compliant way. Furthermore, we have discussed how to visually compare localities, even in a many-criteria scenario. Finally, we have acomplished this analysis building on sound TSA methodology, which have allowed us to (i)~mitigate the impact of seasonality and noise, (ii)~aggregate temporal dynamics using different perspectives, and; (iii)~assess the data using different temporal discretization granularities.

To empirically demonstrate the proposed approach, we have conducted both a region- and a country-level analysis on mobility reduction rates from localities in different continents. Some of the insights observed have been already reported, such as the need for calibration and seasonality approaches, as well as the drawbacks with $\epsilon$-dominance if extreme tolerance values are adopted. Other insights concern CMR data specifically, e.g. the disagreement between rankings as a function of the experimental factors considered. Specifically, we have demonstrated that place categories, aggregation measures, and seasonality approaches interact, and that a coarse temporal discretization conceals dynamics that are critical to SD monitoring.

Though the MCDM methodology discussed here has helped identify interesting insights, follow-up investigations can further leverage these benefits. For instance, future work could bridge multi-criteria TSA methodology with simulation analysis to predict the impact of SD policies. Another important research direction is to explore more visualization techniques to improve the MCDM process for public policymakers. Finally, sociodemographic aspects are likely related to how localities rank, and the examples illustrated in this work could serve as grounds for that investigation.

\bibliographystyle{model1-num-names}

\bibliography{refs}








\end{document}